# Habitability is a continuous property of nature

René Heller
Max Planck Institute for Solar System Research, Justus-von-Liebig-Weg 3, 37077 Göttingen, Germany
heller@mps.mpg.de
14 April 2020

In their recent comment, Cockell et al.[1] argue that the habitability of an environment is fundamentally a binary property; that is to say, an environment can either support the metabolic processes of a given organism or not. The habitability of different environments, they argue, may have different degrees that could be determined at least in theory by answering the question: 'is this environment habitable to a given organism?' 'More' or 'less' habitable environments could then be related by the number of yes or no answers given to what is fundamentally a series of binary questions and decisions.

In my opinion, there are at least three implicit assumptions made for this line of reasoning that are implausible and that sabotage the conclusions. The first is in the genetic diversity of the organisms. The wording of Cockell et al. suggests that the organisms subjected to a hypothetical yes/no inquiry are in fact distinct as, for example, the cyanobacterium *Cylindrospermum* and the domestic pig, *Sus domesticus*. This implicit assumption of a biologically discontinuous sample of organisms is key to the logic of the binary interpretation of habitability.

Let us conduct a gedankenexperiment that invokes a sample of four organisms: three closely related types of cyanobacteria ($c_1$, $c_2$, $c_3$) and a pig (Fig. 1). Now let us consider two environments. Environment 1 (E1) is an Earth-like planet but with relatively nonsaline oceans in which $c_2$ and $c_3$ cannot survive, but $c_1$ can tolerate these conditions and thrive. This planet shall be the habitat of $c_1$ and the pig. Environment 2 (E2) is a highly saline ocean planet much unlike Earth that has no land-based life and only a few different types of single-celled organisms, all of which are cyanobacteria. In this cyanobacteria world, we envision that $c_1$, $c_2$, and $c_3$ can happily prosper, while pigs obviously can't. From the perspective of Cockell et al., E2 would be more habitable than E1 because the number of organisms that can live in it is larger than for E1. This experiment shows that from this perspective, the habitability of an environment depends on the sample, in particular on its composition and size, but it ignores the genetic complexity of the sample. I argue that the genetic diversity of the sample members is one realization from a continuous spectrum of possibilities, for example via random mutations. As a consequence, the sample is fundamentally continuous in nature and a binary construction of the sample is not meaningful. The conclusion of a binary interpretation for habitability must thus be flawed.

The second misconception is in the assumption that an ecosystem can be decomposed into subsets of independent samples. This is equivalent to the notion that life itself is discontinuous in terms of the realizations (organisms) that it generates. In my view, a single cyanobacterium, and probably also a single pig, may be able to live in a given sterile laboratory environment for a limited amount of time, for example to answer the question 'is this environment habitable to this organism?'. But organisms need food. This food must contain living organisms or at least derivatives of dead organisms. A single pig might be kept alive with purely synthetic food for some time, but this would render the definition of habitability into a 'survive until starve' equivalence. As far as we know from the one and only ecosystem that we can study, organisms are links in a food chain or, maybe more adequately, nodes in a food web. This is just one of many reasons why a discontinuous sample of organisms cannot even be produced in the first place. The picture of habitability as an emerging property of what are fundamentally binary states is thus not a realistic description of nature.

The third problem is in the definition of an environment. Such a definition is challenging to construct but for a thorough logical derivation of the binarity/non-binarity concept of habitability, which is to be evaluated for a particular environment, we first must answer the question: 'what is an environment?' Is it the interior of a human skin cell that is the habitat to a bacterium? Or is it the surface of a small rock? Or the shoreline of the Mediterranean Sea? Or is it rather the surface of an entire planet? Does our very own human 'environment' actually include the Moon, which affects the ocean tides and the longterm rotational state of the Earth? What about the sun as our primary source of energy? The answer may even depend on the organism that one is interested in because some aspects are more relevant to one organism while being irrelevant to others. As a consequence, there simply exists no discontinuous concept of an environment and we understand that it must be as continuous as the breadth of organisms, be it in a genetical or phenotypic sense, and so on. The question of the environment is a continuous one in both space and time and thus any concept constructed to be applicable to a sample of environments must be continuous as well. In fact, with Earth scoring 1 on a 0/1 binary scale, a binary concept of habitability obscures the possibility that planets could also be 'more habitable' than Earth[2], be it in terms of their biomass, genetic diversity, longevity of habitability, resilience against asteroidal destruction et cetera.

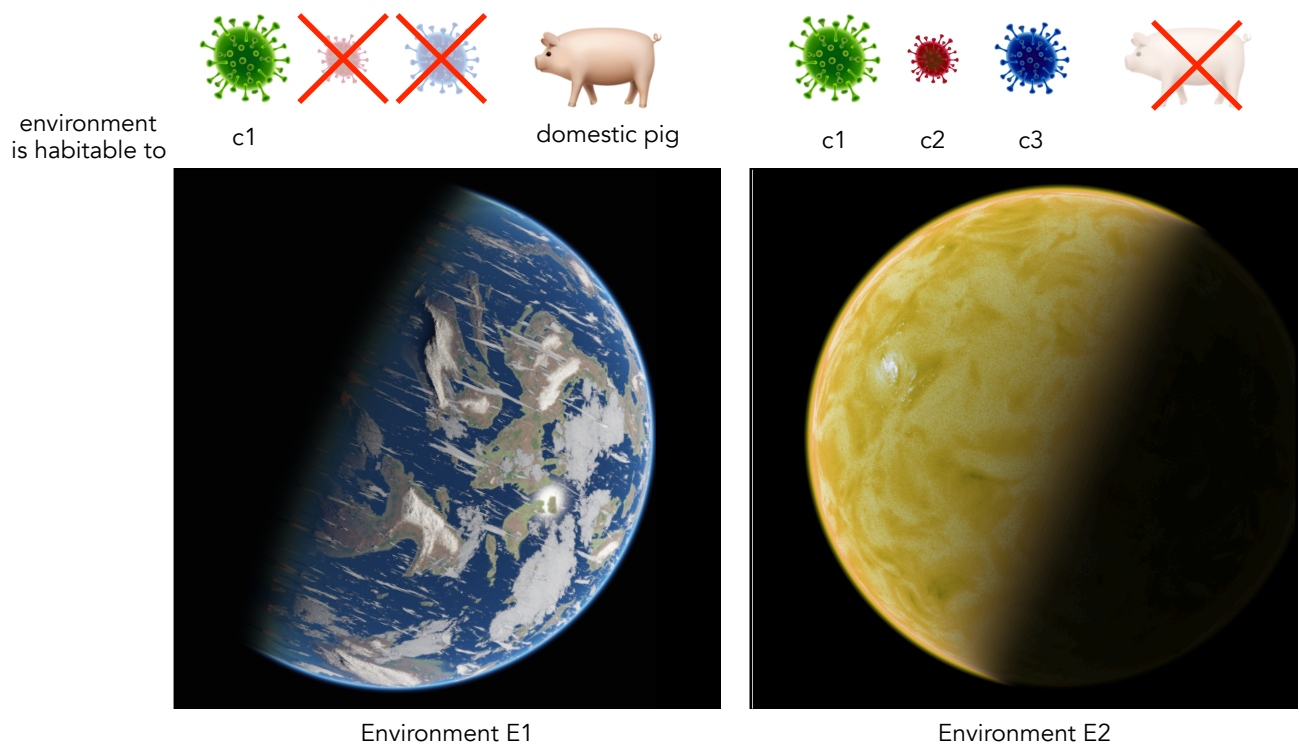

**Fig. 1 | Habitability ranking based on a binary interpretation of habitability.** Consider two environments E1 (left) and E2 (right). E1 is an Earth-like planet that offers habitable conditions for cyanobacterium c1 and pigs, whereas E2 is a highly saline ocean world that only harbors cyanobacteria and that is habitable to c1, c2, and c3. In the binary interpretation of habitability, E2 is more habitable than E1 although it is genetically less diverse.

In summary, we find that the question of whether habitability is a binary quantity or not brings us back to the question of whether our concepts of life and of the environments that life thrives in (or not) are binary or non-binary. I argue that the latter is the case, and hence any modern concept of habitability must be continuous too.

As a final remark, Cockell et al. state that the binarity aspects of habitability can assist us in advancing the clarity in the use of this term. This perspective implies that the word 'habitability' must first be discussed or defined before it can be used properly. But the opposite is true. Linguistic philosophy has long argued that the meaning of a word is given by its usage in language[3]. Hence, the way we speak about habitability defines what we mean with it, and not the other way around. Once we acknowledge that we speak about habitability in the continuous context of life and its environments, then we understand that habitability is continuous.

## Acknowledgements

This work was supported by the German Aerospace Center (DLR) under PLATO Data Center grant 50OL1701.